\documentclass[12pt]{article}
\usepackage{amssymb}
\def\hybrid{\topmargin 0pt      \oddsidemargin 0pt
        \headheight 0pt \headsep 0pt
        \voffset=-0.5cm
        \hoffset=-0.25in
        \textwidth 6.75in
        \textheight 9.5in       % A4 paper
        \marginparwidth 0.0in
        \parskip 5pt plus 1pt   \jot = 1.5ex}
\catcode`\@=11
\def\marginnote#1{}

\newcount\hour
\newcount\minute
\newtoks\amorpm
\hour=\time\divide\hour by60 \minute=\time{\multiply\hour by60
\global\advance\minute by-\hour}
\edef\standardtime{{\ifnum\hour<12 \global\amorpm={am}%
        \else\global\amorpm={pm}\advance\hour by-12 \fi
        \ifnum\hour=0 \hour=12 \fi
        \number\hour:\ifnum\minute<10 0\fi\number\minute\the\amorpm}}
\edef\militarytime{\number\hour:\ifnum\minute<10 0\fi\number\minute}

\def\draftlabel#1{{\@bsphack\if@filesw {\let\thepage\relax
   \xdef\@gtempa{\write\@auxout{\string
      \newlabel{#1}{{\@currentlabel}{\thepage}}}}}\@gtempa
   \if@nobreak \ifvmode\nobreak\fi\fi\fi\@esphack}
        \gdef\@eqnlabel{#1}}
\def\@eqnlabel{}
\def\@vacuum{}
\def\draftmarginnote#1{\marginpar{\raggedright\scriptsize\tt#1}}
\def\draftlabel#1{{\@bsphack\if@filesw {\let\thepage\relax
   \xdef\@gtempa{\write\@auxout{\string
      \newlabel{#1}{{\@currentlabel}{\thepage}}}}}\@gtempa
   \if@nobreak \ifvmode\nobreak\fi\fi\fi\@esphack}
        \gdef\@eqnlabel{#1}}
\def\@eqnlabel{}
\def\@vacuum{}
\def\draftmarginnote#1{\marginpar{\raggedright\scriptsize\tt#1}}

\def\draft{\oddsidemargin -.5truein
        \def\@oddfoot{\sl preliminary draft \hfil
        \rm\thepage\hfil\sl\today\quad\militarytime}
        \let\@evenfoot\@oddfoot \overfullrule 3pt
        \let\label=\draftlabel
        \let\marginnote=\draftmarginnote
   \def\@eqnnum{(\theequation)\rlap{\kern\marginparsep\tt\@eqnlabel}%
\global\let\@eqnlabel\@vacuum}  }

\def\numberbysection{\@addtoreset{equation}{section}
        \def\theequation{\thesection.\arabic{equation}}}

\def\underline#1{\relax\ifmmode\@@underline#1\else
        $\@@underline{\hbox{#1}}$\relax\fi}

\def\titlepage{\@restonecolfalse\if@twocolumn\@restonecoltrue\onecolumn
     \else \newpage \fi \thispagestyle{empty}\c@page\z@
        \def\thefootnote{\fnsymbol{footnote}} }

\def\endtitlepage{\if@restonecol\twocolumn \else  \fi
        \def\thefootnote{\arabic{footnote}}
        \setcounter{footnote}{0}}  %\c@footnote\z@ }
\numberbysection \hybrid

\newcounter{mo}

\newcommand{\ti}[1]{\tilde{#1}}

\newcommand{\vf}{\varphi}
\newcommand{\al}{\alpha}
\newcommand{\be}{\beta}

\newcommand{\om}{\omega}
\newcommand{\vth}{\vartheta}

\newcommand{\Mat}{ {\rm Mat}(N,\mathbb C) }

\newcommand{\mC}{\mathbb C}
\newcommand{\mZ}{\mathbb Z}

\newcommand{\Om}{\Omega}

\newcommand{\z}{{\zeta}}

\newtheorem{predl}{Proposition}[section]

\def\beq{\begin{equation}}
\def\eq{\end{equation}}
\def\p{\partial}

\def\res{\mathop{\hbox{Res}}\limits}

\begin{document}

\setcounter{page}{1}

%\date{}
%\date{}
%\vspace{50mm}

\begin{flushright}
 ITEP-TH-30/19\\
\end{flushright}
\vspace{0mm}

\begin{center}
\vspace{0mm}
 {\LARGE{Odd supersymmetric Kronecker elliptic function}}
\\ \vspace{4mm} {\LARGE{and Yang-Baxter equations}}
%\\ \vspace{4mm} {\LARGE{related $R$-matrices and Yang-Baxter equations}}
% \\ \vspace{4mm} {\LARGE{and associative Yang-Baxter equation}}
% \\ \vspace{4mm}  {\LARGE{and KZB equation}}
\\
\vspace{15mm} {\large \ \ {A. Levin}\,{\small $^{\S\, \ddagger}$}
 \ \ \ \ \ {M. Olshanetsky}\,{\small $^{\ddagger\, \flat\, \natural}$}
 \ \ \ \ \ {A. Zotov\,}%\footnote{The corresponding author. Data available on request from the authors.}
 {\small $^{\diamondsuit\, \ddagger\, \S\, \natural}$} }
 \vspace{10mm}

 \vspace{1mm}$^\S$ - {\small{\rm National Research University Higher School of Economics, Russian Federation,  \\
%  Mathematics Department of
 Usacheva str. 6,  Moscow, 119048, Russia}}
 \\
%
 %\vspace{1mm} $^\ddagger$ -- {\small{\rm %Institute of Theoretical and Experimental Physics, 117218,  Moscow, Russia
 %ITEP, B. Cheremushkinskaya str. 25,  Moscow, 117218, Russia}}
 %
 \vspace{1mm} $^\ddagger$ -- {\small{\rm Institute for Theoretical and Experimental Physics of
  NRC ''Kurchatov Institute'',\\
 B. Cheremushkinskaya str. 25,  Moscow, 117218, Russia}}
 \\
 \vspace{1mm} $^\flat$ --
 {\small{\rm Institute for Information Transmission Problems RAS (Kharkevich Institute),\\
 Bolshoy Karetny per. 19, Moscow, 127994, Russia}}
\\
\vspace{1mm} $^\diamondsuit$ -- {\small{\rm
 Steklov Mathematical Institute of Russian Academy of Sciences,\\ Gubkina str. 8, Moscow,
119991,  Russia}}
 \\
  \vspace{1mm} $^\natural$ -- {\small{\rm Moscow Institute of Physics and Technology,\\ Inststitutskii per.  9, Dolgoprudny,
 Moscow region, 141700, Russia}}

\end{center}

%\vspace{2mm}
\begin{center}\footnotesize{{\rm E-mails:}{\rm\
alevin2@hse.ru,\ olshanet@itep.ru,\ zotov@mi-ras.ru}}\end{center}
%
%\vspace{0mm}
%

 \begin{abstract}
We introduce an odd supersymmetric version of the Kronecker elliptic
function. It satisfies the genus one Fay identity and supersymmetric
version of the heat equation. As an application we construct an odd
supersymmetric extensions of the elliptic $R$-matrices, which
satisfy the classical and the associative Yang-Baxter equations.
 \end{abstract}

\bigskip

\bigskip

%\newpage
%\tableofcontents

\section{Introduction}
\setcounter{equation}{0}

Consider an elliptic curve $\Sigma_\tau={\mC}/(\mZ\oplus\tau\mZ)$
with moduli $\tau$, ${\rm Im}(\tau)>0$. The elliptic Kronecker
function \cite{Weil}
  \beq\label{a01}
  \begin{array}{c}
  \displaystyle{
\phi(\hbar,z;\tau)\equiv\phi(\hbar,z)=\frac{\vth'(0)\vth(\hbar+z)}{\vth(\hbar)\vth(z)}
 }
 \end{array}
 \eq
is defined on $\Sigma_\tau$ in terms of the theta-function
  \beq\label{a02}
  \begin{array}{c}
  \displaystyle{
\vth(z;\tau)\equiv\vth(z)=\displaystyle{\sum _{k\in \mathbb Z}} \exp
\left ( \pi \imath \tau (k+\frac{1}{2})^2 +2\pi \imath
(z+\frac{1}{2})(k+\frac{1}{2})\right )\,,
 }
 \end{array}
 \eq
 which has simple zero at
 $z=0$ due to skew-symmetry $\vth(z)=-\vth(-z)$.
 Therefore, the function (\ref{a01}) has a simple pole at $z=0$
  \beq\label{a06}
  \begin{array}{c}
  \displaystyle{
 \res\limits_{z=0}\phi(\hbar,z)=1\,.
 }
 \end{array}
 \eq
 The quasi-periodic behavior on the lattice $\mZ\oplus\tau\mZ$ is as
 follows:
  \beq\label{a07}
  \begin{array}{c}
  \displaystyle{
 \phi(\hbar,z+1)=\phi(\hbar,z)\,,\quad\quad \phi(\hbar,z+\tau)=e^{-2\pi\imath
 \hbar}\phi(\hbar,z)\,.
 }
 \end{array}
 \eq
These two properties (\ref{a06}), (\ref{a07}) fix the Kronecker
function explicitly as it is given in (\ref{a01}).

 For our purposes the most important property of (\ref{a01}) is that it satisfies the following quadratic
 relation called the \underline{genus one Fay trisecant
 identity \cite{Fay}}:
  \beq\label{a03}
  \begin{array}{c}
  \displaystyle{
\phi(\hbar_1,z_{12})\phi(\hbar_2,z_{23})=\phi(\hbar_2,z_{13})\phi(\hbar_1-\hbar_2,z_{12})+
\phi(\hbar_2-\hbar_1,z_{23})\phi(\hbar_1,z_{13})\,.
 }
 \end{array}
 \eq
 The next important property of the Kronecker function is that it satisfies the \underline{heat equation}:
  \beq\label{a10}
  \begin{array}{c}
  \displaystyle{
 2\pi\imath\p_\tau\phi(\hbar,z;\tau)=\p_z\p_\hbar\phi(\hbar,z;\tau)\,.
 }
 \end{array}
 \eq
 This one follows from the heat equation for the theta-function
 (\ref{a02}): $4\pi\imath\p_\tau\vth(z;\tau)=\p_z^2\vth(z;\tau)$.

Using the skew-symmetry
  \beq\label{a04}
  \begin{array}{c}
  \displaystyle{
\phi(\hbar,z_{12})=-\phi(-\hbar,z_{21})
 }
 \end{array}
 \eq
 rewrite (\ref{a03}) in the form
  \beq\label{a05}
  \begin{array}{c}
  \displaystyle{
\phi(\hbar_1,z_{12})\phi(\hbar_2,z_{23})+\phi(-\hbar_2,z_{31})\phi(\hbar_1-\hbar_2,z_{12})
+\phi(\hbar_2-\hbar_1,z_{23})\phi(-\hbar_1,z_{31})=0\,,
 }
 \end{array}
 \eq
so that the Kronecker function %(\ref{a01})
 is a scalar representation of
the Fomin-Kirillov algebra \cite{FK} of $A_2$ type.
Relation (\ref{a03}) or (\ref{a05}) and its degenerations are widely
used in classical and quantum integrable systems
\cite{KrSkl,Belavin,Kirillov,LOZ}. The main reason is
 that it underlies the Yang-Baxter equations.
 In this paper we deal with two types of the
 equations (see details in Section \ref{sect4}):
 \begin{itemize}
 \item classical Yang-Baxter equation
  \beq\label{a08}
  \displaystyle{
  [r_{12}(z_{12}),r_{13}(z_{13})]+[r_{12}(z_{12}),r_{23}(z_{23})]+[r_{13}(z_{13}),r_{23}(z_{23})]=0\,.
  }
 \eq
 \item associative Yang-Baxter equation \cite{FK,Pol}
  \beq\label{a11}
    \displaystyle{
  R^{\hbar_1}_{12}(z_{12})
 R^{{\hbar_2}}_{23}(z_{23})=R^{{\hbar_2}}_{13}(z_{13})R_{12}^{{\hbar_1}-{\hbar_2}}(z_{12})+
 R^{{\hbar_2}-{\hbar_1}}_{23}(z_{23})R^{\hbar_1}_{13}(z_{13})\,.
 }
  \eq
%
% \item quantum Yang-Baxter equation
% %
%  \beq\label{a09}
%    \displaystyle{
%R_{12}^\hbar(z_{12})R_{13}^\hbar(z_{13})R_{23}^\hbar(z_{23})=
%R_{23}^\hbar(z_{23})R_{13}^\hbar(z_{13})R_{12}^\hbar(z_{12})\,.
% }
%  \eq
%
 \end{itemize}

\noindent{\bf Purpose of the paper} is to construct supersymmetric
generalization of the Kronecker function (\ref{a02}) in such a way
that the Fay identity (\ref{a03}) or (\ref{a05}) remains valid.
The notion of supersymmetric elliptic curve (supertorus with odd
spin structure) together with definitions of supersymmetric version
of elliptic functions was introduced in \cite{Levin} and
\cite{Rabin}. See also \cite{DP} for applications and further
developments. The supersymmetric version of the Kronecker function
can be defined in different ways. For example, one can define it
through the ratio (\ref{a01}) of supersymmetric theta-functions
 proposed in papers \cite{Levin,Rabin}. But the Fay identity is not valid
  for this type generalization. In this paper we suggest an
alternative construction, which solves the problem.

 Let $\z_k,\mu,\om$ be a set of Grassmann variables, that
 is\footnote{$[a,b]_+=ab+ba$ is the anticommutator, while $[a,b]_-=[a,b]=ab-ba$ stands for commutator.}
  \beq\label{a191}
  \begin{array}{c}
  \displaystyle{
 \z_k^2=\mu_i^2=\om^2=0\,,\quad
 [\z_k,\z_l]_+=[\z_k,\mu_i]_+=[\mu_i,\mu_j]_+=[\z_k,\om]_+=[\om,\mu_i]_+=0\,.
 }
 \end{array}
 \eq
Introduce the following odd function:
  \beq\label{a20}
  \begin{array}{c}
  \displaystyle{
 {\bf\Phi}(\hbar,z_1,z_2;\tau |\, \mu,\z_1,\z_2;\om)\equiv {\bf\Phi}^{\hbar|\,\mu}(z_1,z_2|\,\z_1,\z_2)
 =(\z_1-\z_2)\phi(\hbar,z_{12})+
 }
 \\ \ \\
  \displaystyle{
 +\om\p_1\phi(\hbar,z_{12})+2\pi\imath\z_1\z_2\om\p_\tau\phi(\hbar,z_{12})
 +\z_1\z_2\mu\p_1\phi(\hbar,z_{12})+\frac{1}{2}(\z_1+\z_2)\mu\om\p_1^2\phi(\hbar,z_{12})\,,
 }
 \end{array}
 \eq
 where
  \beq\label{a19}
  \begin{array}{c}
  \displaystyle{
\p_1\phi(x,y)=\p_x\phi(x,y)\,,\quad \p_2\phi(x,y)=\p_y\phi(x,y)\,.
 }
 \end{array}
 \eq
 These notations will be used in what follows.
Also, by definition
  \beq\label{a25}
  \begin{array}{c}
  \displaystyle{
 {\bf\Phi}^{\hbar|\,0}(z_1,z_2|\,\z_1,\z_2)
 =(\z_1-\z_2)\phi(\hbar,z_{12})
 +\om\p_1\phi(\hbar,z_{12})+2\pi\imath\z_1\z_2\om\p_\tau\phi(\hbar,z_{12})
 \,,
 }
 \end{array}
 \eq
 which is (\ref{a20}) without two last terms.
 The third term in (\ref{a20}) or (\ref{a25}) can be transformed via (\ref{a10})
as
$2\pi\imath\z_1\z_2\om\p_\tau\phi(\hbar,z_{12})=\z_1\z_2\om\p_1\p_2\phi(\hbar,z_{12})$.
Instead of skew-symmetry (\ref{a04}) we now have the symmetry
property
  \beq\label{a21}
  \begin{array}{c}
  \displaystyle{
{\bf\Phi}^{\hbar|\,\mu}(z_1,z_2|\,\z_1,\z_2)={\bf\Phi}^{-\hbar|\,-\mu}(z_2,z_1|\,\z_2,\z_1)\,.
 }
 \end{array}
 \eq
 We will prove that (\ref{a20}) and (\ref{a25}) satisfy the Fay identity in the
 form
 (\ref{a05}). Namely,
$$
  \displaystyle{
 {\bf\Phi}^{\hbar_1|\,\mu_1}(z_1,z_2|\,\z_1,\z_2){\bf\Phi}^{\hbar_2|\,\mu_2}(z_2,z_3|\,\z_2,\z_3)
 +{\bf\Phi}^{-\hbar_2|\,-\mu_2}(z_3,z_1|\,\z_3,\z_1){\bf\Phi}^{\hbar_1-\hbar_2|\,\mu_1-\mu_2}(z_1,z_2|\,\z_1,\z_2)
 }
$$
  \beq\label{a24}
  \begin{array}{c}
  \displaystyle{
 +{\bf\Phi}^{\hbar_2-\hbar_1|\,\mu_2-\mu_1}(z_2,z_3|\,\z_2,\z_3){\bf\Phi}^{-\hbar_1|\,-\mu_1}(z_3,z_1|\,\z_3,\z_1)=0\,.
 }
 \end{array}
 \eq
 Then we show that (\ref{a20}) satisfies the following relation:
  \beq\label{a30}
  \begin{array}{c}
  \displaystyle{
 \Big( \p_\om+2\pi\imath(\z_1+\z_2)\p_\tau
 \Big){\bf\Phi}^{\hbar|\,\mu}(z_1,z_2|\,\z_1,\z_2)
 =\Big(\p_{\z_1}+\z_1\p_{z_1}-\frac12\mu\p_\hbar\Big)\p_\hbar
 {\bf\Phi}^{\hbar|\,\mu}(z_1,z_2|\,\z_1,\z_2)\,,
 }
 \end{array}
 \eq
which we call the supersymmetric version of the heat equation.

 The paper is organized as follows. In the next Section we derive
 the expression (\ref{a20}) from supersymmetric analogues of the
 simple pole condition (\ref{a06}) and the quasi-periodic boundary
 condition (\ref{a07}). Then we prove the Fay identity (\ref{a24})
 and the odd supersymmetric version (\ref{a30}) of the heat
 equation.
 In Section \ref{sect4} we use the function (\ref{a20}) to construct odd elliptic
 $R$-matrices satisfying supersymmetric versions of the Yang-Baxter
 equations (\ref{a08})-(\ref{a11}). A summary of results is given in
 the Conclusion.

%Levin HSE ???????????

% Russian Academy of Sciences program ''Nonlinear dynamics: fundamental
%problems and applications''.
%The research of A. Zotov was supported
%in part by the HSE University Basic Research Program, Russian
% Academic Excellence Project '5-100' and by the Young Russian Mathematics award.

\section{Supersymmetric Kronecker function}\label{sect2}
\setcounter{equation}{0}

In order to construct supersymmetric generalization of the Kronecker
function let us specify its definition in the ordinary case. As
mentioned previously, it is fixed as the ratio of theta-functions
(\ref{a01}) by two conditions: to have a simple pole at $z=0$ with
residue (\ref{a06}) and by the quasi-periodic boundary conditions
(\ref{a07}). Equivalently, the function $\phi(\hbar,z_1-z_2)$ is
fixed as the Green function $\phi(\hbar,z_1,z_2)$ of
$\bar\p=\p_{\bar{z}_1}$-operator
  \beq\label{a51}
  \begin{array}{c}
  \displaystyle{
  \res\limits_{z_1=z_2}\phi(\hbar,z_1,z_2)=1\quad\hbox{or}\quad
  \bar\p\phi(\hbar,z_1,z_2)=1
 }
 \end{array}
 \eq
with the boundary conditions
  \beq\label{a071}
  \begin{array}{c}
  \displaystyle{
 \phi(\hbar,z_1+1,z_2)=\phi(\hbar,z_1,z_2)\,,\quad\quad \phi(\hbar,z_1+\tau,z_2)=e^{-2\pi\imath
 \hbar}\phi(\hbar,z_1,z_2)\,,
 }
 \\ \ \\
   \displaystyle{
\phi(\hbar,z_1,z_2+1)=\phi(\hbar,z_1,z_2)\,,\quad\quad
\phi(\hbar,z_1,z_2+\tau)=e^{2\pi\imath
 \hbar}\phi(\hbar,z_1,z_2)\,.
  }
 \end{array}
 \eq
 Then the solution of (\ref{a51})-(\ref{a071}) is given by
 $\phi(\hbar,z_1,z_2)=\phi(\hbar,z_1-z_2)$.

\paragraph{Odd supersymmetric Kronecker elliptic function.}
While an elliptic curve $\Sigma_\tau$ with moduli $\tau$ is a
quotient of $\mC$ (with a coordinate $z$) by translations
$z\rightarrow z+1$, $z\rightarrow z+\tau$, the corresponding super
elliptic curve ${\bf\Sigma}_{\tau,\om}$ is a quotient of $\mC^{1|1}$
(with coordinates $z,\z$) by (super)translations
  \beq\label{a33}
  \left\{
  \begin{array}{l}
 z\rightarrow z+1\,,
 \\
 \z\rightarrow \z\,,
 \end{array}
 \right.
  \qquad\qquad
  \left\{
  \begin{array}{l}
 z\rightarrow z+\tau+2\pi\imath\z\om\,,
 \\
 \z\rightarrow \z+2\pi\imath\om\,.
 \end{array}
 \right.
 \eq
 It is equipped with the
covariant derivative $D_\z=\p_\z+\z\p_z$, $D_\z^2=\p_z$. In what
follow we use the Grassmann variables $\z_k,\om,\mu_i$ as
superpartners to the coordinates $z_k$, to the moduli $\tau$ and to
 $\mC$-valued parameters of the boundary conditions
 $\hbar_i$ respectively:
% \footnote{The
%conformal weights of the variables are as follows. For $z_k,\hbar_i$
%it equals 1, for $\tau$ it equals 2. For the odd variables
%$\z_k,\mu_k$ the conformal weights are equal to $1/2$, and the one
%for $\om$ is equal to $3/2$. }
%:
%
  \beq\label{a31}
  \begin{array}{cccc}
 \hbox{even variables:}\ &\  z_k\  &\  \tau\ &\  \hbar_i\
 \\
  \hbox{odd variables:}\ &\  \z_k\  &\ \om\  &\  \mu_i\
 \end{array}
 \eq

Similarly to (\ref{a51})-(\ref{a071}) the supersymmetric Kronecker
function (\ref{a20}) is defined on a product of two super elliptic
curves. It is fixed by the following two conditions:
 \begin{itemize}
 \item it has a simple pole on a diagonal $z_1=z_2$ with residue
  \beq\label{a341}
  \begin{array}{c}
  \displaystyle{
  \res\limits_{z_{1}=z_2}{\bf\Phi}^{\hbar|\,\mu}(z_1,z_2|\,\z_1,\z_2)=\z_1-\z_2\,.
 }
 \end{array}
 \eq

 \item it is a
quasi-periodic function with respect to (super)translations
(\ref{a33}):
  \beq\label{a34}
  \begin{array}{l}
  \displaystyle{
 {\bf\Phi}^{\hbar|\,\mu}(z_1+1,z_2|\,\z_1,\z_2)={\bf\Phi}^{\hbar|\,\mu}(z_1,z_2+1|\,\z_1,\z_2)
 ={\bf\Phi}^{\hbar|\,\mu}(z_1,z_2|\,\z_1,\z_2)\,,
 }
 \\ \ \\
  \displaystyle{
 {\bf\Phi}^{\hbar|\,\mu}(z_1+\tau+2\pi\imath\z_1\om,z_2|\,\z_1+2\pi\imath\om,\z_2)=
 { g}_1{\bf\Phi}^{\hbar|\,\mu}(z_1,z_2|\,\z_1,\z_2)\,,
 }
  \\ \ \\
  \displaystyle{
 {\bf\Phi}^{\hbar|\,\mu}(z_1,z_2+\tau+2\pi\imath\z_2\om|\,\z_1,\z_2+2\pi\imath\om)=
 { g}_2{\bf\Phi}^{\hbar|\,\mu}(z_1,z_2|\,\z_1,\z_2)\,,
 }
 \end{array}
 \eq
where
  \beq\label{a35}
  \begin{array}{c}
  \displaystyle{
  { g}_1=\exp\Big( -2\pi\imath
  (\hbar+\mu\z_1+\pi\imath\mu\om)
  \Big)\,,
  \quad
 { g}_2=\exp\Big( 2\pi\imath
  (\hbar+\mu\z_2-\pi\imath\mu\om)
  \Big)\,.
 }
 \end{array}
 \eq
 \end{itemize}
%
%??? Notice that the boundary conditions depend on $\z_1$ and do not
%depend on $\z_2$.
 So that ${\bf\Phi}^{\hbar|\,\mu}(z_1,z_2|\,\z_1,\z_2)$ is the Green
 function of the $\bar\p=\p_{{\bar z}_1}$ operator with the boundary
 conditions (\ref{a34}).

 \begin{predl}
 The boundary conditions (\ref{a34})-(\ref{a35}) holds true for the function
 (\ref{a20}).
 For the truncated Kronecker function (\ref{a25}) the transition function (\ref{a35}) is given by
  $g=\exp(-2\pi\imath\hbar)$.
\end{predl}
\noindent\underline{\em{Proof:}}\quad Let us prove the statement for
the truncated function. Consider the  first term in (\ref{a25}).
Under translations (\ref{a33}) it transforms as
  \beq\label{a53}
  \begin{array}{c}
  \displaystyle{
  (\z_1-\z_2)\phi(\hbar,z_{12})\rightarrow
  (\z_1-\z_2+2\pi\imath\om)\phi(\hbar,z_{12}+\tau+2\pi\imath\z_1\om)=
 }
 \\ \ \\
  \displaystyle{
=(\z_1-\z_2+2\pi\imath\om)\Big(\phi(\hbar,z_{12}+\tau)+2\pi\imath\z_1\om\p_2\phi(\hbar,z_{12}+\tau)\Big)=
  }
   \\ \ \\
  \displaystyle{
=\exp(-2\pi\imath\hbar)\Big((\z_1-\z_2)\phi(\hbar,z_{12})+2\pi\imath\om\phi(\hbar,z_{12})
+2\pi\imath\z_1\z_2\om\phi(\hbar,z_{12})\Big)\,.
  }
 \end{array}
 \eq
 Therefore, the first term in (\ref{a20}) is not quasi-periodic. As a result of (super)translation it acquires
 additional unwanted terms proportional to $\om\phi(\hbar,z_{12})$ and
 $\z_1\z_2\om\phi(\hbar,z_{12})$. They are compensated by the
 contributions coming from the second and the third terms of (\ref{a20}). Indeed,
 using (\ref{a071}) one can easily verify that
  \beq\label{a54}
  \begin{array}{c}
  \displaystyle{
 \om\p_1\phi(\hbar,z_{12})\rightarrow
 \om\exp(-2\pi\imath\hbar)\Big( \p_1\phi(\hbar,z_{12})-2\pi\imath\phi(\hbar,z_{12})
 \Big)\,,
 }
 \\ \ \\
  \displaystyle{
 \z_1\z_2\om\p_\tau\phi(\hbar,z_{12})\rightarrow
 \z_1\z_2\om\exp(-2\pi\imath\hbar)\Big( \p_\tau\phi(\hbar,z_{12})-\p_2\phi(\hbar,z_{12})
 \Big)\,.
  }
 \end{array}
 \eq
 Then, summing up (\ref{a53})-(\ref{a54}) we conclude that the
 truncated function is quasi-periodic with the multiplicator
 $\exp(-2\pi\imath\hbar)$.
The rest of the proof for the function (\ref{a20}) also uses
  \beq\label{a55}
  \begin{array}{c}
  \displaystyle{
 \p_1^2\phi(\hbar,z_{12}+\tau)=
 \exp(-2\pi\imath\hbar)\Big(
 \p_1^2\phi(\hbar,z_{12})-4\pi\imath\p_1\phi(\hbar,z_{12})-4\pi^2\phi(\hbar,z_{12})
 \Big)\,.
 }
 \end{array}
 \eq
 The calculations are performed in a similar way. $\blacksquare$

Conversely, one can derive  (\ref{a20}), (\ref{a25}) from conditions
(\ref{a341})-(\ref{a35}). For example, to reproduce the truncated
function (\ref{a25}) from (\ref{a341})-(\ref{a35}) with
$g=\exp(-2\pi\imath\hbar)$ one should start with the first term in
(\ref{a25}). Its presence in the final expression follows from
(\ref{a341}) and (\ref{a07}). Under the translations (\ref{a33}) is
transformed as given in (\ref{a53}). In order to compensate the
unwanted terms one should use the terms from (\ref{a54}). A more
general answer (\ref{a20}) is reproduced in the same way.

\paragraph{Fay identity.}
Let us prove the key property of the supersymmetric Kronecker
function.
 \begin{predl}
 The Fay identity (\ref{a24}) holds true for the function (\ref{a20}).
\end{predl}
\noindent\underline{\em{Proof:}}\quad The verification of the
statement is a tedious but straightforward calculation. One should
substitute the definition (\ref{a20}) into (\ref{a24}) and write
down the coefficients behind all possible monomials of (distinct)
Grassmann variables. For example, the coefficient behind monomials
$\z_1\z_2$, $\z_2\z_3$ and $\z_3\z_1$ is given by (the l.h.s. of)
the ordinary Fay identity (\ref{a05}). The coefficients behind
$\z_1\om$, $\z_2\om$ and $\z_3\om$ are obtained from (\ref{a05}) by
the action of operators $\p_{\hbar_2}$, $\p_{\hbar_1}+\p_{\hbar_2}$
and $\p_{\hbar_1}$ respectively. All other coefficients are also
identities, which follows from the Fay identity (\ref{a05}) by
taking some derivatives. $\blacksquare$

 Let us slightly rewrite the definition of the Kronecker function (\ref{a20})
 using the heat equation (\ref{a10}) for its third term:
  \beq\label{a55011}
  \begin{array}{c}
  \displaystyle{
  {\bf\Phi}(\hbar,z_1,z_2;\tau |\, \mu,\z_1,\z_2;\om)=
  }
 \\ \ \\
  \displaystyle{
  =\Big[(\z_1-\z_2)
 +\om\p_\hbar+\z_1\z_2\om\p_\hbar\p_{z_1}
 +\z_1\z_2\mu\p_\hbar+\frac{1}{2}(\z_1+\z_2)\mu\om\p_\hbar^2\Big]\phi(\hbar,z_1-z_2)\,.
 }
 \end{array}
 \eq
This formula can be applied to the rational and trigonometric
degenerations, where the moduli $\tau$ is absent. For example, in
the trigonometric case the
 function (\ref{a01}) turns into
 $\phi(\hbar,z_{12})=\coth(\hbar)+\coth(z_{12})$. Then (\ref{a55011})
 is equal to
  \beq\label{a55041}
  \begin{array}{c}
  \displaystyle{
 {\bf\Phi}^{\rm trig}=
 (\z_1-\z_2)\Big( \coth(\hbar)+\coth(z_{1}-z_2)
 \Big)
  -\frac{\om+\z_1\z_2\mu}{\sinh^2(\hbar)}+\frac{(\z_1+\z_2)\mu\om\cosh(\hbar)}{\sinh^3(\hbar)}\,.
 }
 \end{array}
 \eq
 Similarly, for the rational degeneration
 $\phi(\hbar,z_{12})=1/\hbar+1/z_{12}$ we have
  \beq\label{a55042}
  \begin{array}{c}
  \displaystyle{
 {\bf\Phi}^{\rm rat}=
 (\z_1-\z_2)\Big( \frac{1}{\hbar}+\frac{1}{z_1-z_2}
 \Big)
  -\frac{\om+\z_1\z_2\mu}{\hbar^2}+\frac{(\z_1+\z_2)\mu\om}{\hbar^3}\,.
 }
 \end{array}
 \eq
 The functions (\ref{a55041}) and (\ref{a55042}) also satisfy the Fay identity (\ref{a24})
 since the functions $\coth(\hbar)+\coth(z_{12})$, $1/\hbar+1/z_{12}$ satisfy (\ref{a05}).

\paragraph{Heat equation.}
%\label{sect3}
%\setcounter{equation}{0}
%
The result is as follows.
 \begin{predl}
 The function $ {\bf\Phi}(\hbar,z_1,z_2;\tau |\, \mu,\z_1,\z_2;\om)$ (\ref{a20})
 satisfies the odd supersymmetric heat equation (\ref{a30}).
  Similarly, for the function (\ref{a25}) we have
  \beq\label{a301}
  \begin{array}{c}
  \displaystyle{
 \Big( \p_\om+2\pi\imath(\z_1+\z_2)\p_\tau
 \Big){\bf\Phi}^{\hbar|\,0}(z_1,z_2|\,\z_1,\z_2)
 =\Big(\p_{\z_1}+\z_1\p_{z_1}\Big)\p_\hbar
 {\bf\Phi}^{\hbar|\,0}(z_1,z_2|\,\z_1,\z_2)\,.
 }
 \end{array}
 \eq
\end{predl}
The proof is straightforward. It uses the ordinary heat equation
(\ref{a10}) only.

%%%%%%%%%%%%%%%%%%%%%%%%%%%%%%%%%%%%%%%%%%%%%%%%%%%%%%%%%%%%%%%%%%%%%%%%%%%%%%%%%%%%%%%%

\section{Yang-Baxter equations}\label{sect4}
\setcounter{equation}{0}

The construction of elliptic $R$-matrix uses special basis in $\Mat$
\cite{Belavin}. The pair  of matrices $Q,\Lambda\in\Mat$
 \beq\label{a71}
 \begin{array}{c}
  \displaystyle{
Q_{kl}=\delta_{kl}\exp\left(\frac{2\pi
 \imath}{N}k\right)\,,\ \ \ \Lambda_{kl}=\delta_{k-l+1=0\,{\hbox{\tiny{mod}}}
 N}\,,\quad Q^N=\Lambda^N=1_{N}
 }
 \end{array}
 \eq
provides the finite-dimensional representation of the Heisenberg
group due to
 \beq\label{a72}
 \begin{array}{c}
  \displaystyle{
 \exp\left(\frac{2\pi\imath}{N}\,a_1
 a_2\right)Q^{a_1}\Lambda^{a_2}=\Lambda^{a_2}Q^{a_1}\,,\quad
 a_1,a_2\in\mZ\,.
 }
 \end{array}
 \eq
 Then the basis in $\Mat$ is given by the following set of $N^2$ matrices:
 \beq\label{a73}
 \begin{array}{c}
  \displaystyle{
 T_a=T_{a_1 a_2}=\exp\left(\frac{\pi\imath}{N}\,a_1
 a_2\right)Q^{a_1}\Lambda^{a_2}\,,\quad
 a=(a_1,a_2)\in\mZ_N\times\mZ_N\,.
 }
 \end{array}
 \eq
 From (\ref{a72}) we have
  \beq\label{a74}
 \begin{array}{c}
  \displaystyle{
T_\al T_\be=\kappa_{\al,\be} T_{\al+\be}\,,\ \ \
\kappa_{\al,\be}=\exp\left(\frac{\pi \imath}{N}(\be_1
\al_2-\be_2\al_1)\right)\,,
 }
 \end{array}
 \eq
 where $\al+\be=(\al_1+\be_1,\al_2+\be_2)$.
% The non-degenerate
% pairing is given by the matrix trace:
%
%  \beq\label{e906}
% \begin{array}{c}
%  \displaystyle{
%\tr(T_\al T_\be)=N\delta_{\al+\be}\,,\quad T_0=1_N\,.
% }
% \end{array}
% \eq
%
 Next, define the set of $N^2$ basis functions numerated by the index
$a=(a_1,a_2)\in\mZ_N\times\mZ_N$:
 \beq\label{a76}
 \begin{array}{c}
  \displaystyle{
 \vf_a(\hbar+\Omega_a,z)=\exp(2\pi\imath\frac{a_2}{N}\,z)\,\phi(\hbar+\Omega_a,z)\,,\quad
 \Omega_a=\frac{a_1+a_2\tau}{N}\,.
 }
 \end{array}
 \eq
 Finally, the quantum Baxter-Belavin's elliptic $R$-matrix is of the form:
 \beq\label{a77}
 \begin{array}{c}
  \displaystyle{
R_{12}^\hbar(z)=\sum\limits_\al T_\al\otimes T_{-\al}\,
 \vf_a(\hbar+\Omega_a,z)\,.
  }
 \end{array}
 \eq
It satisfies the quantum Yang-Baxter equation and the associative
Yang-Baxter equation (\ref{a11}) \cite{Pol}. Similarly, the
classical Belavin-Drinfeld-Sklyanin $r$-matrix
 \beq\label{a78}
 \begin{array}{c}
  \displaystyle{
 r_{12}(z)=\sum\limits_{\al\neq 0} T_\al\otimes T_{-\al}\,
 \vf_a(\Omega_a,z)
  }
 \end{array}
 \eq
 satisfies the classical Yang-Baxter equation (\ref{a08}), which
  is based on the Fay identity (\ref{a03}) or (\ref{a05}) written as relations
  for the functions (\ref{a76}):
  \beq\label{a79}
  \begin{array}{c}
  \displaystyle{
 \vf_\al(\Om_\al,z_{12})\vf_\be\Om_\be,z_{23})
 +\vf_{-\be}(-\Om_\be,z_{31})\vf_{\al-\be}(\Om_{\al-\be},z_{12})+
  }
 \\ \ \\
 \displaystyle{
+\vf_{\be-\al}(\Om_{\be-\al},z_{23})\vf_{-\al}(-\Om_\al,z_{31})=0\,,\quad
 \al,\be,\al-\be\neq 0\,.
 }
 \end{array}
 \eq
 In the same way the associative Yang-Baxter equation (\ref{a11}) is
 based on a more general relation
  \beq\label{a60}
  \begin{array}{c}
  \displaystyle{
 \vf_\al(\hbar+\Om_\al,z_{12})\vf_\be(\eta+\Om_\be,z_{23})
 +\vf_{-\be}(-\eta-\Om_\be,z_{31})\vf_{\al-\be}(\hbar-\eta+\Om_{\al-\be},z_{12})+
 }
 \\ \ \\
 \displaystyle{
+\vf_{\be-\al}(\eta-\hbar+\Om_{\be-\al},z_{23})\vf_{-\al}(-\hbar-\Om_\al,z_{31})=0\,.
 }
 \end{array}
 \eq

\paragraph{Supersymmetric basis functions.} In order to construct
supersymmetric generalizations
 of (\ref{a77}) and  (\ref{a78}) we
need an analogue of the basis functions (\ref{a76}). They have the
following form:
  \beq\label{a80}
 \begin{array}{c}
  \displaystyle{
 {\bf\Phi}_\al^{\hbar+\Omega_\al|\,\mu}(z_1,z_2|\,z_1,\z_2):=
 \exp\Big( 2\pi\imath\frac{\al_2}{N}(z_1-z_2+\z_1\z_2)
 \Big){\bf\Phi}^{\hbar+\Omega_\al|\,\mu}(z_1,z_2|\,z_1,\z_2)=
 }
 \\ \ \\
  \displaystyle{
=\Big(1+2\pi\imath\frac{\al_2}{N}\z_1\z_2\Big){\bf\Phi}^{\hbar+\Omega_\al|\,\mu}(z_1,z_2|\,z_1,\z_2)=
 }
 \\ \ \\
  \displaystyle{
=
{\bf\Phi}^{\hbar+\Omega_\al|\,\mu}(z_1,z_2|\,z_1,\z_2)+2\pi\imath\frac{\al_2}{N}\z_1\z_2\om\p_1\phi(\hbar+\Omega_\al,z_{12})\,.
  }
 \end{array}
 \eq
 Equivalently, the set of functions is written in the forms
  \beq\label{a81}
 \begin{array}{c}
  \displaystyle{
 {\bf\Phi}_\al^{\hbar+\Omega_\al|\,\mu}(z_1,z_2|\,z_1,\z_2)=
 \exp\Big( 2\pi\imath\frac{\al_2}{N}(z_1-z_2) \Big){\bf\Phi}^{\hbar+\Omega_\al|\,\mu+2\pi\imath\frac{\al_2}{N}\om}(z_1,z_2|\,z_1,\z_2)
 }
 \end{array}
 \eq
 or
  \beq\label{a82}
 \begin{array}{c}
  \displaystyle{
 {\bf\Phi}_\al^{\hbar+\Omega_\al|\,\mu}(z_1,z_2|\,z_1,\z_2)=
 \exp\Big( 2\pi\imath\frac{\al_2}{N}(z_1-z_2)
 \Big)\ti{\bf\Phi}_\al^{\hbar+\Omega_\al|\,\mu}(z_1,z_2|\,z_1,\z_2)\,,
 }
 \end{array}
 \eq
where
  \beq\label{a83}
  \begin{array}{c}
  \displaystyle{
 \ti{\bf\Phi}_\al^{\hbar+\Omega_\al|\,\mu}(z_1,z_2|\,z_1,\z_2)
 =(\z_1-\z_2)\vf_\al(\hbar+\Omega_\al,z_{12})+
 \om\p_1\vf_\al(\hbar+\Omega_\al,z_{12})+
 }
 \\ \ \\
  \displaystyle{
+2\pi\imath\z_1\z_2\om\frac{d}{d\tau}\vf_\al(\hbar+\Omega_\al,z_{12})
 +\z_1\z_2\mu\p_1\vf_\al(\hbar+\Omega_\al,z_{12})+
 }
 \\ \ \\
  \displaystyle{
 +\frac{1}{2}(\z_1+\z_2)\mu\om\p_1^2\vf_\al(\hbar+\Omega_\al,z_{12})\,.
 }
 \end{array}
 \eq
 In the third term of (\ref{a83}) the full derivative with respect to $\tau$ includes also
 the partial derivative with respect to the first argument of $\vf_\al(\hbar+\Omega_\al,z_{12})$, depending
 on $\tau$ through $\Om_\al$ (\ref{a76}).

Using the Fay identity (\ref{a24}) it is easy to show that the set
of functions (\ref{a80}) satisfy the following direct analogue of
(\ref{a60}):
  \beq\label{a84}
  \begin{array}{c}
  \displaystyle{
 {\bf\Phi}_{\al}^{\hbar_1+\Omega_\al|\,\mu_1}(z_1,z_2|\,\z_1,\z_2){\bf\Phi}_\be^{\hbar_2+\Omega_\be|\,\mu_2}(z_2,z_3|\,\z_2,\z_3)+
 }
  \\ \ \\
  \displaystyle{
 +{\bf\Phi}_{-\be}^{-\hbar_2-\Omega_\be|\,-\mu_2}(z_3,z_1|
 \,\z_3,\z_1){\bf\Phi}_{\al-\be}^{\hbar_1-\hbar_2+\Omega_{\al-\be}|\,\mu_1-\mu_2}(z_1,z_2|\,\z_1,\z_2)+
 }
 \\ \ \\
  \displaystyle{
 +{\bf\Phi}_{\be-\al}^{\hbar_2-\hbar_1+\Omega_{\be-\al}|\,\mu_2-\mu_1}(z_2,z_3|
 \,\z_2,\z_3){\bf\Phi}_{-\al}^{-\hbar_1-\Omega_\al|\,-\mu_1}(z_3,z_1|\,\z_3,\z_1)=0\,.
 }
 \end{array}
 \eq
 In the same way for $\al,\be,\al-\be\neq (0,0)$ we also have
  \beq\label{a841}
  \begin{array}{c}
  \displaystyle{
 {\bf\Phi}_{\al}^{\Omega_\al|\,0}(z_1,z_2|\,\z_1,\z_2){\bf\Phi}_\be^{\Omega_\be|\,0}(z_2,z_3|\,\z_2,\z_3)
 +{\bf\Phi}_{-\be}^{-\Omega_\be|\,0}(z_3,z_1|
 \,\z_3,\z_1){\bf\Phi}_{\al-\be}^{\Omega_{\al-\be}|\,0}(z_1,z_2|\,\z_1,\z_2)+
 }
 \\ \ \\
  \displaystyle{
 +{\bf\Phi}_{\be-\al}^{\Omega_{\be-\al}|\,0}(z_2,z_3|
 \,\z_2,\z_3){\bf\Phi}_{-\al}^{-\Omega_\al|\,0}(z_3,z_1|\,\z_3,\z_1)=0\,.
 }
 \end{array}
 \eq

\paragraph{Classical super Yang-Baxter equation.}
 The odd supersymmetric analog of the classical $r$-matrix
 (\ref{a78}) is as follows:
  \beq\label{a90}
  \begin{array}{c}
  \displaystyle{
{\bf r}_{12}(z_1,z_2|\,\z_1,\z_2)=\sum\limits_{\al\neq 0}
T_\al\otimes T_{-\al}
{\bf\Phi}_{\al}^{\Omega_\al|\,0}(z_1,z_2|\,\z_1,\z_2)\,.
 }
 \end{array}
 \eq
%
% We deal with the odd functions, which anticommute.
For the odd
 $r$-matrix the
 classical Yang-Baxter equations were studied in
  \cite{Kulish,Kirillov}\footnote{Let us remark that we do not consider super Lie algebras (or groups)
  as it is discussed in \cite{Kulish}.
   We deal with ${\rm GL}_N$ $R$-matrices in fundamental representation.}.
   The super version of the equation
 (\ref{a08}) contains anticommutators instead of commutators:
  \beq\label{a91}
  \begin{array}{c}
  \displaystyle{
 [{\bf r}_{12},{\bf r}_{13}]_+
 +
 [{\bf r}_{12},{\bf r}_{23}]_+
 +
[{\bf r}_{13},{\bf r}_{23}]_+=0\,.
 }
 \end{array}
 \eq
 The following statement holds true.
 \begin{predl}
 The odd supersymmetric analog of the classical $r$-matrix
 (\ref{a90}) satisfies equation (\ref{a91}), where ${\bf r}_{ab}={\bf
 r}_{ab}(z_a,z_b|\,\z_a,\z_b)$.
 \end{predl}
 The proof is similar to the one given below for a more general
 $R$-matrix.

\paragraph{Associative Yang-Baxter-equation.} Using the skew-symmetry
 $R_{12}^\hbar(z)=-R_{21}^{-\hbar}(-z)$ of (\ref{a77}) let us rewrite
equation (\ref{a11}) in the form
  \beq\label{a111}
    \displaystyle{
  R^{\hbar_1}_{12}(z_{12})
 R^{\hbar_2}_{23}(z_{23})+R^{-\hbar_2}_{31}(z_{31})R_{12}^{\hbar_1-\hbar_2}(z_{12})+
 R^{\hbar_2-\hbar_1}_{23}(z_{23})R^{-\hbar_1}_{31}(z_{31})=0\,,
 }
  \eq
  which is similar to (\ref{a05}).
 The odd supersymmetric analog of the quantum  elliptic Baxter-Belavin $R$-matrix
 (\ref{a77}) is as follows:
  \beq\label{a40}
  \begin{array}{c}
  \displaystyle{
 {\bf R}_{12}^{\hbar|\,\mu}(z_1,z_2|\,\z_1,\z_2)=\sum\limits_{\al} T_\al\otimes T_{-\al}
 {\bf\Phi}_{\al}^{\hbar+\Omega_\al|\,\mu}(z_1,z_2|\,\z_1,\z_2)\,.
 }
 \end{array}
 \eq
 \begin{predl}
 The odd supersymmetric analog of the quantum $R$-matrix
 (\ref{a40}) satisfies the following equation:
  \beq\label{a41}
  \begin{array}{c}
  \displaystyle{
{\bf R}_{12}^{\hbar_1|\,\mu_1}{\bf R}^{\hbar_2|\,\mu_2}_{23}
 +{\bf R}_{31}^{-\hbar_2|\,-\mu_2}{\bf R}_{12}^{\hbar_1-\hbar_2|\,\mu_1-\mu_2}
 +{\bf R}_{23}^{\hbar_2-\hbar_1|\,\mu_2-\mu_1}{\bf R}_{31}^{-\hbar_1|\,-\mu_1}=0\,.
 }
 \end{array}
 \eq
 where ${\bf R}_{ab}^{\hbar|\,\mu}={\bf
 R}_{ab}^{\hbar|\,\mu}(z_a,z_b|\,\z_a,\z_b)$.
 \end{predl}
\noindent\underline{\em{Proof:}}\quad The proof is similar to the
ordinary case. One should multiply the l.h.s. of the identity
(\ref{a84}) by $\Mat^{\otimes 3}$ valued element $\kappa_{\be,\al}
T_{\al}\otimes T_{\be-\al}\otimes T_{-\be}$, and then sum up over
indices $\al,\be\in\mZ_N\times\mZ_N$. In order to prove it let us
write down the first term from the l.h.s. of (\ref{a41}). Using
(\ref{a74}) we have
 $$
 {\bf R}_{12}^{\hbar_1|\,\mu_1}{\bf R}^{\hbar_2|\,\mu_2}_{23}=
 $$
  \beq\label{a42}
  \begin{array}{c}
  \displaystyle{
 \sum\limits_{\al,\be}\kappa_{-\al,\be} T_{\al}\otimes T_{\be-\al}\otimes T_{-\be}
 {\bf\Phi}_{\al}^{\hbar_1+\Omega_\al|\,\mu_1}(z_1,z_2|\,\z_1,\z_2){\bf\Phi}_\be^{\hbar_2+\Omega_\be|\,\mu_2}(z_2,z_3|\,\z_2,\z_3)\,.
 }
 \end{array}
 \eq
 The second term from the l.h.s. of (\ref{a41}) has the form:
 $$
 {\bf R}_{31}^{-\hbar_2|\,-\mu_2}{\bf R}_{12}^{\hbar_1-\hbar_2|\,\mu_1-\mu_2}=
 $$
  \beq\label{a43}
  \begin{array}{c}
  \displaystyle{
 \sum\limits_{\al,\be}\kappa_{\be,\al-\be} T_{\al}\otimes T_{\be-\al}\otimes T_{-\be}
 {\bf\Phi}_{-\be}^{-\hbar_2-\Omega_\be|\,-\mu_2}(z_3,z_1|
 \,\z_3,\z_1){\bf\Phi}_{\al-\be}^{\hbar_1-\hbar_2+\Omega_{\al-\be}|\,\mu_1-\mu_2}(z_1,z_2|\,\z_1,\z_2)\,.
 }
 \end{array}
 \eq
 And the third term from the l.h.s. of (\ref{a41}) is of the form:
 $$
 {\bf R}_{23}^{\hbar_2-\hbar_1|\,\mu_2-\mu_1}{\bf R}_{31}^{-\hbar_1|\,-\mu_1}=
 $$
  \beq\label{a44}
  \begin{array}{c}
  \displaystyle{
 \sum\limits_{\al,\be}\kappa_{\al-\be,-\al} T_{\al}\otimes T_{\be-\al}\otimes T_{-\be}
 {\bf\Phi}_{\be-\al}^{\hbar_2-\hbar_1+\Omega_{\be-\al}|\,\mu_2-\mu_1}(z_2,z_3|
 \,\z_2,\z_3){\bf\Phi}_{-\al}^{-\hbar_1-\Omega_\al|\,-\mu_1}(z_3,z_1|\,\z_3,\z_1)\,.
 }
 \end{array}
 \eq
 The statement of the Proposition then follows from
 $\kappa_{\be,\al}=\kappa_{-\al,\be}=\kappa_{\be,\al-\be}=\kappa_{\al-\be,-\al}$.
 The latter comes from (\ref{a74}). $\blacksquare$

%%%%%%%%%%%%%%%%%%%%%%%%%%%%%%%%%%%%%%%%%%%%%%%%%%%%%%%%%%%%%%%%%%%%%%%%%%%%%%%%%%%%%%%%%%%%%%%%%%%%%%

\section{Conclusion}
\setcounter{equation}{0}

We introduced the odd supersymmetric version of the elliptic
Kronecker function (\ref{a01}):
  \beq\label{a5501}
  \begin{array}{c}
  \displaystyle{
  {\bf\Phi}(\hbar,z_1,z_2;\tau |\, \mu,\z_1,\z_2;\om)=
  }
 \\ \ \\
  \displaystyle{
  =\Big[(\z_1-\z_2)
 +\om\p_\hbar+2\pi\imath\z_1\z_2\om\p_\tau
 +\z_1\z_2\mu\p_\hbar+\frac{1}{2}(\z_1+\z_2)\mu\om\p_\hbar^2\Big]\phi(\hbar,z_1-z_2)\,.
 }
 \end{array}
 \eq
 It satisfies the Fay identity (\ref{a24}) and the
 supersymmetric version of the heat equation (\ref{a30}). Both
 equations also hold true for the truncated function (\ref{a25}). In
 this case one should replace $\mu$ with $0$ in (\ref{a24}), (\ref{a30}).

 Using (\ref{a5501}) we constructed supersymmetric extension of the
 elliptic $R$-matrix (\ref{a40}).
 It follows from (\ref{a83}) that similarly to (\ref{a5501})  it can be represented in the
 form:
  \beq\label{a5502}
  \begin{array}{c}
  \displaystyle{
 {\bf  R}_{12}^{\hbar|\,\mu}(z_1,z_2|\,\z_1,\z_2)=
  }
 \\ \ \\
  \displaystyle{
  =\Big[(\z_1-\z_2)
 +\om\p_\hbar+2\pi\imath\z_1\z_2\om\p_\tau
 +\z_1\z_2\mu\p_\hbar+\frac{1}{2}(\z_1+\z_2)\mu\om\p_\hbar^2\Big]R_{12}^{\hbar}(z_{12})\,.
 }
 \end{array}
 \eq
In fact, (\ref{a5502}) contains (\ref{a5501}) as particular case
 (when $N=1$). By changing the third term via the heat equation
(\ref{a10}) we get
  \beq\label{a5503}
  \begin{array}{c}
  \displaystyle{
 {\bf  R}_{12}^{\hbar|\,\mu}(z_1,z_2|\,\z_1,\z_2)=
  }
 \\ \ \\
  \displaystyle{
  =\Big[(\z_1-\z_2)
 +\om\p_\hbar+\z_1\z_2\om\p_\hbar\p_{z_1}
 +\z_1\z_2\mu\p_\hbar+\frac{1}{2}(\z_1+\z_2)\mu\om\p_\hbar^2\Big]R_{12}^{\hbar}(z_{12})\,,
 }
 \end{array}
 \eq
 which is applicable in trigonometric and rational cases
 corresponding to nodal and cuspidal degenerations of the elliptic
 curve.

Finally, the $R$-matrix (\ref{a5503}) was proved to satisfy the
associative Yang-Baxter equation written as
  \beq\label{a5505}
  \begin{array}{c}
  \displaystyle{
{\bf R}_{12}^{\hbar_1|\,\mu_1}{\bf R}^{\hbar_2|\,\mu_2}_{23}
 +{\bf R}_{31}^{-\hbar_2|\,-\mu_2}{\bf R}_{12}^{\hbar_1-\hbar_2|\,\mu_1-\mu_2}
 +{\bf R}_{23}^{\hbar_2-\hbar_1|\,\mu_2-\mu_1}{\bf R}_{31}^{-\hbar_1|\,-\mu_1}=0\,.
 }
 \end{array}
 \eq
 In the same way the supersymmetric version of the classical
 $r$-matrix (\ref{a90}) was shown to solve the classical super
 Yang-Baxter equation (\ref{a91}).

 The supersymmetric heat equation together with the Fay identity can
 be used for construction of the Knizhnik-Zamolodchikov-Bernard
 equations or the quantum Schlesinger system on supersymmetric elliptic curves. We will describe it in
 our next paper.

% Conflict of Interest: The authors declare that they have no conflicts of interest.

%%%%%%%%%%%%%%%%%%%%%%%%%%%%%%%%%%%%%%%%%%%%%%%%%%%%%%%%%%%%%%%%%%%%%%%%%%%%%%%%%%%%%%%%%%%%%%%%%%%%%%

\paragraph{Acknowledgments.} The work
was supported in part by RFBR grants 18-02-01081 (A. Levin and M.
Olshanetsky), 18-01-00926 (A. Zotov) and by joint RFBR project
19-51-18006 Bolg$_a$ (M. Olshanetsky).
% The work of A. Levin was
%partially supported by Laboratory of Mirror Symmetry NRU HSE, RF
%Government grant, ag. 14.641.31.0001.
The research of A. Zotov was also supported in part by the HSE University Basic Research Program, Russian Academic Excellence Project '5-100' and by the Young Russian Mathematics award.
% The research of A. Zotov was also supported in part by the Young Russian Mathematics award.

\begin{small}
 
\end{small}


\begin{thebibliography}{99}
\addcontentsline{toc}{section}{References}

%\footnotesize{

\bibitem{Belavin}
R.J. Baxter, %{\em Partition function of the eight-vertex lattice
%model},
Ann. Phys. 70 (1972) 193--228.
\\
 A.A. Belavin,
% {\em Dynamical symmetry of integrable quantum systems},
Nucl. Phys. B, 180 (1981) 189--200.
% \\
% M.P. Richey, C.A. Tracy,
%{\em  ${\mathbb Z}_n$ Baxter model: Symmetries and the Belavin
%parametrization }
%J. Stat. Phys. 42 (1986) 311--348.
%
%\\
%A.A. Belavin, V.G. Drinfeld,
% {\em Solutions of the classical Yang–Baxter equation for simple Lie algebras},
%Funct. Anal. Appl., 16:3 (1982) 159–-180.
%\\
%E.K. Sklyanin, %{\em Some algebraic structures connected with the Yang—-Baxter equation},
%Funct. Anal. Appl., 16:4 (1982) 263--270.

\bibitem{DP} E.D'Hoker, D.H. Phong,
% Conformal scalar fields and chiral splitting on super Riemann surfaces
Commun. Math. Phys. 125 (1989) 469--513.
\\
A. Rogers, {\em Supermanifolds Theory and Applications}, World
Scientific, 2007.
\\
E. Witten,
% {\em Notes on super Riemann surfaces and their moduli},
Pure and Applied Mathematics Quarterly, 15:1 (2019) 57--211,
% Special Issue on Super Riemann Surfaces and String Theory;
\\  arXiv:1209.2459 [hep-th].


\bibitem{Fay} J.D. Fay,
 {\em Theta Functions on Riemann Surfaces}, Lecture Notes in Mathematics, Vol. 352 (1973) Springer-Verlag Berlin.





\bibitem{FK} S. Fomin, A.N. Kirillov,
%{\em Quadratic algebras, Dunkl elements, and Schubert calculus},
Advances in geometry; Progress in Mathematics book series, Vol. 172
 (1999) 147--182.


\bibitem{Kirillov}
A.N. Kirillov,
 % {\em On Some Quadratic Algebras I 12  :
 %Combinatorics of Dunkl and Gaudin Elements, Schubert, Grothendieck,
 %Fuss-Catalan, Universal Tutte and Reduced Polynomials},
 SIGMA 12 (2016), 002;  arXiv:1502.00426 [math.RT].

%\bibitem{Mumford}  D. Mumford, {\em Tata Lectures on Theta I, II},
%Birkh\"auser, Boston, Mass. (1983, 1984).


\bibitem{KrSkl}
I.M.  Krichever, %{ Elliptic solutions of the Kadomtsev--Petviashvili
%equation and integrable systems of particles},
Funct.  Anal.  Appl., 14:4 (1980) 282--290.
\\
E.K. Sklyanin, %{\em Some algebraic structures connected with the Yang—-Baxter equation},
Funct. Anal. Appl., 16:4 (1982) 263--270.
\\
E.K. Sklyanin, %{\em Some algebraic structures connected with the
%Yang—-Baxter equation. Representations of quantum algebras},
Funct. Anal. Appl., 17:4 (1983) 273--284.



\bibitem{Kulish} P.P. Kulish, E.K. Sklyanin,
%
 J. Soviet Math., 19:5 (1982) 1596--1620
 \\
S.M. Khoroshkin, V.N. Tolstoy, Commun. Math. Phys., 141:3 (1991)
599--617.






\bibitem{Levin} A. M. Levin, %"Supersymmetric elliptic curves",
Funct. Anal. Appl., 21:3 (1987) 243--244.
\\
A. M. Levin, %"Supersymmetric elliptic and modular functions",
Funct. Anal. Appl., 22:1 (1988) 60--61.


\bibitem{LOZ} A. Levin, M. Olshanetsky, A. Zotov,
%{\em Planck Constant as
%Spectral Parameter in Integrable Systems and KZB Equations},
JHEP 10 (2014) 109; arXiv:1408.6246 [hep-th].
\\
A.M. Levin, M.A. Olshanetsky, A.V. Zotov,
%{\em Quantum Baxter-Belavin R-matrices and multidimensional Lax pairs for Painleve VI},
Theoret. and Math. Phys. 184:1 (2015) 924--939;\\
 arXiv:1501.07351 [math-ph].
 \\
 A.V. Zotov,
 % "Calogero–Moser model and R -matrix identities",
  Theoret. and Math. Phys., 197:3 (2018),
1755--1770.

%\bibitem{MM} Yu.I. Manin, S.A. Merkulov,
%{\em Semisimple Frobenius (super)manifolds and quantum cohomology of Pr}
%Topological Methods in Nonlinear Analysis,
%Vol. 9, Num. 1 (1997) 107--161.

 \bibitem{Pol} A. Polishchuk, % {\em Classical Yang–Baxter equation and the $A^\infty$-constraint},
Advances in Mathematics 168:1 (2002)  56--95.


\bibitem{Rabin} J.M. Rabin, P.G.O. Freund,
% "Supertori are algebraic curves,"
Commun. Math. Phys. 114 (1988) 131--145.
\\
J.M. Rabin,
% Super Elliptic Curves
    J. Geom. and Phys. 15 (1995) 252--280;     arXiv:hep-th/9302105.

\bibitem{Weil} A. Weil, {\em Elliptic functions according to Eisenstein and
Kronecker}, Springer-Verlag, %Berlin- Heidelberg-New York,
 (1976).
 \\
 D. Mumford, {\em Tata Lectures on Theta I, II},
Birkh\"auser, Boston, Mass. (1983, 1984).



%\bibitem{Witten} E. Witten,
% {\em Notes on super Riemann surfaces and their moduli},
%Pure and Applied Mathematics Quarterly, 15:1 (2019) 57--211, Special
%Issue on Super Riemann Surfaces and String Theory;  arXiv:1209.2459
%[hep-th].

%}

\end{thebibliography}
\end{document}